\documentclass[11pt]{article}
\usepackage[margin=1.0in]{geometry}
\usepackage{amsmath,amssymb,graphicx}
\usepackage{hyperref}
\usepackage{slashed}
\usepackage[utf8]{inputenc}
\usepackage{pdflscape}
\usepackage{upgreek}
\usepackage{booktabs}
\usepackage{breqn}
\usepackage[font=small,labelfont=bf]{caption}
\usepackage{cite}
\usepackage{enumerate}
\usepackage[table]{xcolor}
\usepackage{makecell}
\usepackage{bm}
\definecolor{gray}{cmyk}{0,0,0,0.05}
\usepackage{multirow}
\usepackage{changepage}
\usepackage{subfigure}

\title{\bf  Correlated Electric and Magnetic Dipole Moment in Natural Supersymmetry}
\author{Sibo Zheng \\
{Department of Physics, Chongqing University, Chongqing 401331, China}}

\begin{document}
\maketitle

\begin{abstract}
In the context of supersymmetry,
the two-loop Barr-Zee diagrams which induce CP-violating electric dipole moment of electron due to superpartners
simultaneously yield CP-conserving magnetic dipole moment of muon. 
In this paper, we derive the coherence between the electric and magnetic dipole moments 
at two-loop level due to stops, charginos or neutralinos-charginos.
We also use the coherence to constrain superpartner masses and their CP-violating phases,
in the light of recent ACME limit on the electric dipole moment of electron and future experiments about
magnetic dipole moment of muon such as Fermilab E989 experiment.
\end{abstract}

\section{Introduction}
In the context of new physics, 
which attempts to address the origin of Higgs, the dark matter with mass of order weak scale,
the puzzle of CP violation, and even the grand unification of standard model (SM) gauge coupling constants,
natural supersymmetry (SUSY) \cite{1302.6587} is one of theoretically well motivated scenarios.
In the minimal version of natural SUSY, there are quite a few sfermions and gauginos with masses of order $\sim 1$ TeV,
which once earned this scenario a promising prospect at experimental facilities such as LHC.
However, the LHC Run I and II have yet observed any new physics except a SM Higgs \cite{Higgs},
which leaves us the only chance at HL-LHC \cite{HL-LHC}.

While waiting for the era of HL-LHC, we may examine other experimental results, especially the latest ones.
Recently, the ACME Collaboration \cite{ACMEII} has updated the measurement on the electric dipole moment (EDM) of electron $d_e$ in 
 \begin{eqnarray}{\label{edm}}
\mathcal{L}_{\text{EDM}}=-i\frac{d_f}{2}\bar{f}\sigma_{\rho\lambda}\gamma^{5}f F^{\rho\lambda},
\end{eqnarray}
which is about an order improvement over the earlier limit \cite{ACMEI}. 
The implications of the new limit to natural SUSY were discussed in \cite{recent}.
For earlier discussions on EDM in the SM and beyond SM, see e.g. \cite{EDMSM} and \cite{EDMReview}, respectively. 
These studies reveal that indirect detections such as the ACME limit on $d_e$ can actually reach mass range of several TeVs, 
which impose stronger constraint on the natural SUSY than the direct detection at LHC. 
In comparison with direct detection,
the main drawback in such indirect detection is that it usually sets a bound on multiple SUSY mass parameters, 
which reduce the prediction.

The situation changes when the same SUSY particle yields multiple independent observables.
As is well known, the one-loop effect \cite{MDM1LOOP} due to light slepton masses \cite{g-2:slepton} 
can explain the observed anomaly of magnetic dipole moment (MDM) of $\mu$ at Brookhaven E821 experiment \cite{0811.1207}.
When the stops \cite{stop}, charginos and neutralinos \cite{electroweakino} running
in the two-loop Barr-Zee diagram \cite{BZ} contribute to $d_e$,
they simultaneously deliver correlated magnetic dipole moment (MDM) of muon $a_{\mu}$ (for a review on MDM of muon, see e.g. \cite{0902.3360}) in,
 \begin{eqnarray}{\label{mdm}}
\mathcal{L}_{\text{MDM}}=\frac{e a_f}{4m_{f}}\bar{f}\sigma_{\rho\lambda}f F^{\rho\lambda}.
\end{eqnarray}
In the literature, earlier discussions on SUSY correlation between $d_e$ and $a_{\mu}$ can be found in \cite{CPtog2},
and two-loop account of $a_{\mu}$ without attention to the coherence in various contexts were presented in \cite{MDM2LOOP}.

In the light of future improvement on the measurements of $d_e$ such as ACME-III limit \cite{ACME-III} and of $a_{\mu}$ 
such as the Fermilab E989 experiment \cite{E989},
the correlation between $d_e$ and $a_{\mu}$ offers a complementary way to constrain the same SUSY particle in the inner loop,
which motivates the materials as follows.
In Sec.2,  we present the theoretic framework for the calculation of two-loop $d_e$ and $a_{\mu}$.
Sec.3 is devoted to explore the numerical analysis of natural SUSY by utilizing the analytic results of Sec.2.
Finally, we conclude in Sec.4.

\section{Correlated EDM and MDM}
The two-loop Barr-Zee diagram can be effectively divided into an inner and an outer loop.
In the inner loop, various SUSY particles such as stops, charginos or neutralinos-charginos run therein.
In the outer loop, there are different mediators such as SM Higgs scalar $h$ or gauge boson $V$. 
For calculating $d_e$, CP-violation is required, and chirality flip always occurs in the Feynman diagram,
which implies that the mediators in the outer loop are identified as one scalar $S=\{h,A\}$ etc 
and a vector boson $V=\{\gamma, Z\}$ or alternatively two $W$ bosons. 
Furthermore, the external photon in either Eq.(\ref{edm}) or Eq.(\ref{mdm}) is only attached to the inner loop,
which together with the mediators ($S$, $V$) constitutes an effective vertex in the outer loop.

What is of interest is that the effective vertex induced by the SUSY particle(s) in the inner loop always
contains both CP-violating and CP-conserving parts. 
In this sense,  the CP-violating contributions to $d_e$ are tied to the CP-conserving ones to $a_{\mu}$.
The coherence provides us a new window to constrain the same SUSY particle(s) in the inner loop.
In what follows, we will consider the effective vertex $\Gamma_{\gamma ...}$ arises from a neutral scalar $S$, a charged scalar $S$ and two $W$ bosons, which will be applied to the cases of stops, charginos, and neutralino-charginos respectively in the next section. 

\subsection{Neutral Scalar}
Firstly, we consider the outer loop composed of a neutral scalar $S$, gauge boson $V$ and photon $\gamma$.
The vertex $\Gamma_{\gamma SV}$ which is induced by superpartner running in the inner loop takes the form at the leading order of photon momentum $q$ \cite{1612.08090}: 
\begin{eqnarray}{\label{vertex}}
\Gamma_{\gamma SV}^{\mu\nu}(q,k)=\int^{1}_{0}dx \frac{1}{x(1-x)}\frac{1}{k^{2}-\Delta(x)}\left[c_{E}(q^{\nu}k^{\mu}-g^{\mu\nu}q\cdot k)+c_{O}\epsilon^{\mu\nu\lambda\rho}q_{\lambda}k_{\rho}\right] 
\end{eqnarray}
where the CP-odd (even) coefficient $c_{O}$ ($c_{E}$) is uniquely determined by the masses and couplings of superpartners running in the inner loop, 
and $\Delta(x)$ in general depends on $x$ and superpartner mass parameters but not $q$ and $k$.

Given the notation on interaction between scalar $S$, vector boson $V$ and SM lepton $f=\{e,\mu\}$ in the outer loop as 
\begin{eqnarray}{\label{notation1}}
\mathcal{L}\supset V_{\mu}\bar{f}(g^{V}_{f}\gamma^{\mu}+g^{A}_{f}\gamma^{\mu}\gamma^{5})f+S\bar{f}(g^{S}_{f}+ig^{P}_{f}\gamma^{5})f
\end{eqnarray}
where all of coupling $g^{S,P}$ and $g^{V,A}$ are real numbers,
we obtain the contribution to EDM of $e$ and MDM of $\mu$ from the same inner loop:
\begin{eqnarray}{\label{analytic1}}
d_{e}&=& -\frac{1}{16\pi^{2}m^{2}_{S}} \int^{1}_{0} dx \frac{1}{x(1-x)}j\left(\frac{m^{2}_{V}}{m^{2}_{S}}, \frac{\Delta}{m^{2}_{S}}\right)\cdot \text{Re}(g_{e}^{V}(c_{O}g_{e}^{S}-c_{E}g_{e}^{P})),\nonumber\\
a_{\mu}&=&-\left(\frac{e}{2m_{\mu}}\right)^{-1}\frac{1}{16\pi^{2}m^{2}_{S}} \int^{1}_{0} dx \frac{1}{x(1-x)}j\left(\frac{m^{2}_{V}}{m^{2}_{S}}, \frac{\Delta}{m^{2}_{S}}\right)\cdot \text{Re}(g_{\mu}^{V}(c_{O}g_{\mu}^{P}+c_{E}g_{\mu}^{S})),
\end{eqnarray}
with function $j(r,s)$ given as,
\begin{eqnarray}{\label{j1}}
j(r,s)=\frac{1}{r-s}\left(\frac{r\log{r}}{r-1}-\frac{s\log{s}}{s-1}\right).
\end{eqnarray}

\subsection{Charged Scalar}
For a complex scalar $S$, the interaction is defined as
\begin{eqnarray}{\label{notation2}}
\mathcal{L}\supset V_{\mu}\bar{f}(g^{V}_{f}\gamma^{\mu}+g^{A}_{f}\gamma^{\mu}\gamma^{5})f'+S\bar{f}(g^{S}_{f}+ig^{P}_{f}\gamma^{5})f'+\text{H.c},
\end{eqnarray}
where couplings $g^{V,A}$ and $g^{S,P}$ are not real in general. 
In this situation, the contribution to EDM of $e$ and MDM of $\mu$ from the same inner loop is given by respectively,
\begin{eqnarray}{\label{analytic2}}
d_{e}&=& -\frac{1}{16\pi^{2}m^{2}_{S}} \int^{1}_{0} dx \frac{1}{x(1-x)}j\left(\frac{m^{2}_{V}}{m^{2}_{S}}, \frac{\Delta}{m^{2}_{S}}\right)\nonumber\\
&\times& [\text{Re}(c_{O}g^{V}_{e}g_{e}^{S*})+\text{Im}(c_{O}g_{e}^{A}g^{P*}_{e})-\text{Re}(c_{E}g^{V}_{e}g_{e}^{P*})+\text{Im}(c_{E}g^{A}_{e}g^{S*}_{e})],\nonumber\\
a_{\mu}&=&-\left(\frac{e}{2m_{\mu}}\right)^{-1}\frac{1}{16\pi^{2}m^{2}_{S}} \int^{1}_{0} dx \frac{1}{x(1-x)}j\left(\frac{m^{2}_{V}}{m^{2}_{S}}, \frac{\Delta}{m^{2}_{S}}\right)\nonumber\\
&\times& [\text{Re}(c_{O}g^{V}_{\mu}g_{\mu}^{P*})-\text{Im}(c_{O}g_{\mu}^{A}g^{S*}_{\mu})+\text{Re}(c_{E}g^{V}_{\mu}g_{\mu}^{S*})+\text{Im}(c_{E}g^{A}_{\mu}g^{P*}_{\mu})],
\end{eqnarray}

\subsection{W Boson Pair}
Unlike in Eq.(\ref{vertex}) the vertex for $\gamma W^{+}W^{-}$ is of form \cite{ww}
\begin{eqnarray}{\label{notation3}}
\Gamma_{\gamma WW}^{\mu\nu\rho}(q,k)=\int^{1}_{0}dx \frac{1}{x(1-x)}\frac{c_{O}}{k^{2}-\Delta(x)}\epsilon^{\mu\nu\rho\sigma}q_{\sigma}
\end{eqnarray}
In this case the contribution to EDM of electron and MDM of moun from the same inner loop described by $c_{O}$ and $\Delta$ reads as respectively, 
\begin{eqnarray}{\label{analytic3}}
d_{e}&=&\frac{1}{16\pi^{2}m^{2}_{W}}\frac{m_{e}}{m^{2}_{W}} \int^{1}_{0} dx \frac{1}{x(1-x)}j\left(\frac{m^{2}_{f'}}{m^{2}_{W}}, \frac{\Delta}{m^{2}_{W}}\right)
\times\text{Im}(c_{O})(\mid g_{\mu}^{V}\mid^{2}+\mid g_{\mu}^{A}\mid^{2}),\nonumber\\
a_{\mu}&=&-\left(\frac{e}{2m_{\mu}}\right)^{-1} \frac{1}{16\pi^{2}m^{2}_{W}}\frac{m_{\mu}}{m^{2}_{W}} \int^{1}_{0} dx \frac{1}{x(1-x)}j\left(\frac{m^{2}_{f'}}{m^{2}_{W}}, \frac{\Delta}{m^{2}_{W}}\right)\times\text{Re}(c_{O})( g_{e}^{V}g_{e}^{A*}+ g_{e}^{A}g_{e}^{V*}) \nonumber\\
\end{eqnarray}

\section{Implications to Natural Supersymmetry}
In the case of stop inner loop, 
the intermediator scalar $S$ in the outer loop can be either the CP-even SM Higgs $h$ or CP-odd scalar $A$.
As previously shown in \cite{1612.08090}, the stop inner loop yields 
$c_{O}=0$ and $c_{E}=\frac{3e^{2}}{4\pi^{2}}Q^{2}_{\tilde{t}_{i}}g^{S}_{\tilde{t}_{i}\tilde{t}_{i}}x(1-x)$,
which implies the intermediator scalar is $h$ ($A$) in the estimate of $a_{\mu}$ ($d_{e}$).
Substituting $g^{V}_{f}=e$, $g^{A}_{f}=0$, $g^{S=h}_{f}= m_{f}/\upsilon$ ($\upsilon=$ 246 GeV), 
and $g^{P=A}_{f}=\tan\beta m_{f}/\upsilon$  into Eq.(\ref{analytic1}), we obtain the 
the dominant contribution to $a_{\mu}$  and $d_{e}$ from the same stop inner loop,
\begin{eqnarray}{\label{stop1}}
\frac{d_{e}/e}{1.1\times 10^{-29}\text{cm}}&=&1.79\times 10^{4}\left(\frac{\tan\beta}{10}\right)\left(\frac{100~\text{GeV}}{m_{A}}\right)^{2}
\sum_{i}\int^{1}_{0} dx(1-x)j(0, \frac{m^{2}_{\tilde{t}_{i}}}{m^{2}_{A}}\frac{1}{x(1-x)})\frac{\text{Re}(g^{S=A}_{\tilde{t}_{i}\tilde{t}_{i}})}{\upsilon}
\nonumber\\
\frac{a_{\mu}}{43\times10^{-10}}&=&- 6.52\times 10^{-3}\sum_{i}\int^{1}_{0} dx(1-x)j(0, \frac{m^{2}_{\tilde{t}_{i}}}{m^{2}_{h}}\frac{1}{x(1-x)}),
\end{eqnarray}
where in the small mixing effect
\begin{eqnarray}{\label{phasestop}}
\frac{\text{Re}(g^{S=A}_{\tilde{t}_{i}\tilde{t}_{i}})}{\upsilon}\simeq
(-1)^{i+1} y^{2}_{t}\frac{\mid A_{t}\mu\mid}{m^{2}_{\tilde{t}_{1}}-m^{2}_{\tilde{t}_{2}}}\sin(\text{arg}(A_{t}\mu))
\end{eqnarray}

\begin{figure}
\centering
\begin{minipage}{8cm}
\centering
\includegraphics[width=8cm,height=4cm]{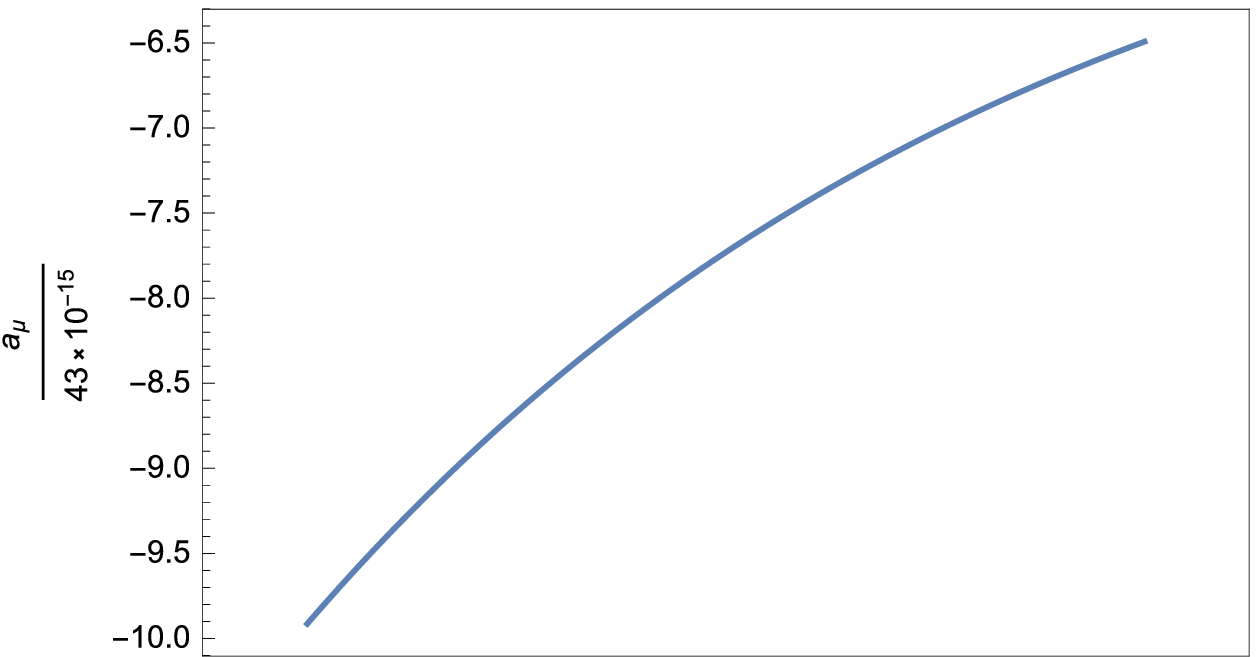}\\
\end{minipage}\\
~~~\centering
\begin{minipage}{8cm}
\centering
\includegraphics[width=7.85cm,height=4cm]{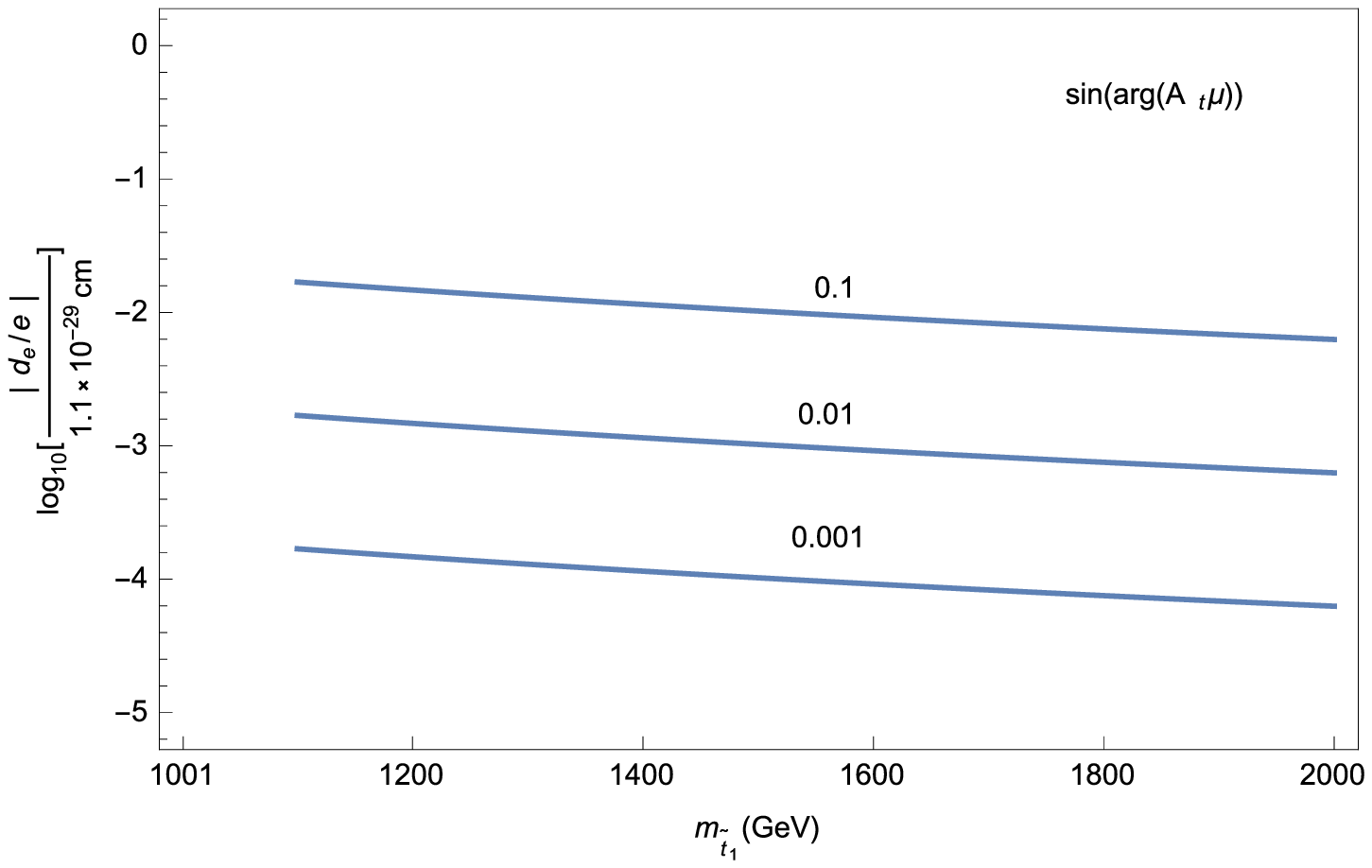}\\
\end{minipage}
 \caption{Correlated $a_{\mu}$ and $d_{e}$ from stop scalars.
 {\it Top}: the dependence of $a_{\mu}$ on stop scalar mass $m_{\tilde{t}_{2}}$ for $m_{\tilde{t}_{1}}=1$ TeV.
 {\it Bottom}: the dependence of $a_{\mu}$ on stop scalar mass $m_{\tilde{t}_{2}}$ for  $\tan\beta=10$, $m_{\tilde{t}_{1}}=m_{A}=1$ TeV, 
 $\mid A_{t}\mu\mid =(100~\text{GeV})^{2}$ and different values of $\sin(\text{arg} (A_{t}\mu))=\{0.1,0.01,0.001\}$.}
\label{stop}
\end{figure}

Fig.\ref{stop} shows the correlation between $a_{\mu}$ and $d_{e}$ from the TeV-scale stops. 
In the {\it top} plot one finds that $a_{\mu}$ is typically of order $10^{-13}$ for stop masses $m_{\tilde{t}_{1}}=1$ TeV and $m_{\tilde{t}_{2}}$ between $1$ TeV and $2$ TeV,
which indicates that the observed anomaly at the Brookhaven E821 experiment has no relevance on stop scalars.
Meanwhile, in the {\it bottom} plot, we observe that the dependence of $d_{e}/e$ (in unit of $1.1\times 10^{-29}$ cm) 
on heavier stop mass $m_{\tilde{t}_{2}}$ for $\tan\beta=10$, small mixing $\mid A_{t}\mu\mid =(100~\text{GeV})^{2}$ 
and different magnitudes of CP-violating phase $\sin(\text{arg} (A_{t}\mu))=\{0.1,0.01,0.001\}$.
This plot shows that the magnitude of $d_{e}/e$ is about two orders below the ACME-II limit. 
In contrast, it can excess the ACME-II limit in the case of large mixing effect and large CP violation \cite{recent} in the stop sector.
Because there is an enhancement on $d_{e}/e$ about two orders of magnitude for large $A_{t}\sim 1$ TeV and large CP-violating phase 
$\sin(\text{arg} (A_{t}\mu))\sim 1$ as shown by Eq.(\ref{phasestop}).
Therefore,  the ACME-II limit imposes strong constraint on stop masses when the mixing effect is large but still weak if the mixing effect is small.

There seems contradiction on the predictions of $A_{t}$ and $\tan\beta$ among different experimental bounds on natural SUSY. 
Firstly, both large mixing and large $\tan\beta$ are required by the observed Higgs mass in the context of MSSM.
Conversely, the ACME experiment strongly favors small $A_t$ and small $\tan\beta$ \cite{bs:susy},
similar to the flavor violation experiment from $b\rightarrow s\gamma$ \cite{bsexp}.
Here we mention a simple way to reconcile the situation.
The Higgs mass can be uplifted by new sources \cite{VLHiggs} such as in the MSSM extended by $5+\bar{5}$ vectorlike fermions \cite{VL} with mass of order TeV scale, which retain the grand unification of SM gauge coupling constants similar to MSSM.
In these scenarios, the stop masses can be of order $\sim 1$ TeV as in natural SUSY, and the mixing effect and $\tan\beta$ can be small.

Unlike the stop inner loop, neither $c_O$ nor $c_E$ in Eq.(\ref{analytic2}) vanishes in the situation of chargino inner loop \cite{1612.08090}.
In this case, the intermediator scalar $S$ can be either $h$ or $A$.
Substituting  $c_{O}=-\frac{e^{2}}{2\pi^{2}}g^{P=A}_{\tilde{\chi}^{+}_{i}\tilde{\chi}^{-}_{j}}(1-x)m_{\tilde{\chi}^{+}}$ and $c_{E}=\frac{e^{2}}{2\pi^{2}}g^{S=h}_{\tilde{\chi}^{+}_{i}\tilde{\chi}^{-}_{j}}(1-x)(2x^{2}-2x+1)m_{\tilde{\chi}^{+}}$ \cite{1612.08090} into Eq.(\ref{analytic2})
results in the EDM of electron and MDM of muon as follows,
\begin{eqnarray}{\label{chargino}}
\frac{d_{e}/e}{1.1\times 10^{-29}\text{cm}}&=&1.09\times 10^{3}
\sum_{i=1,2}\frac{m_{\tilde{\chi}_{i}^{+}}}{100~\text{GeV}}\int^{1}_{0} \frac{dx}{x}\nonumber\\ 
&\times&\left[0.64 j(0, \frac{\Delta}{m^{2}_{h}})\text{Re}g^{P=h}_{\tilde{\chi}^{+}_{i}\tilde{\chi}^{-}_{j}}
+(2x^{2}-2x+1)\tan\beta\left(\frac{100~\text{GeV}}{m_{A}}\right)^{2}j(0, \frac{\Delta}{m^{2}_{A}})\text{Re}g^{S=A}_{\tilde{\chi}^{+}_{i}\tilde{\chi}^{-}_{j}}\right]\nonumber\\
\frac{a_{\mu}}{43\times10^{-10}}&=&6.2 \times 10^{-3}
\sum_{i=1,2}\frac{m_{\tilde{\chi}_{i}^{+}}}{100~\text{GeV}}\int^{1}_{0} \frac{dx}{x}\nonumber\\ 
&\times&\left[0.64(2x^{2}-2x+1) j(0, \frac{\Delta}{m^{2}_{h}})\text{Re}g^{S=h}_{\tilde{\chi}^{+}_{i}\tilde{\chi}^{-}_{j}}
+\tan\beta\left(\frac{100~\text{GeV}}{m_{A}}\right)^{2}j(0, \frac{\Delta}{m^{2}_{A}})\text{Re}g^{P=A}_{\tilde{\chi}^{+}_{i}\tilde{\chi}^{-}_{j}}\right]\nonumber\\\end{eqnarray}
where $\Delta=m^{2}_{\tilde{\chi}_{i}^{+}}/x(1-x)$; 
the CP-violating Yukawa couplings in $d_e$ are given by
$\text{Re}(g^{P=h}_{\tilde{\chi}^{+}_{i}\tilde{\chi}^{-}_{j}})\simeq\frac{g}{\sqrt{2}}\sin\beta\mid C^{R*}_{1i}C^{L}_{2j}\mid\sin (\text{arg}(M_{2}\mu))$ and
$\text{Re}(g^{S=A}_{\tilde{\chi}^{+}_{i}\tilde{\chi}^{-}_{j}})\simeq-\frac{g}{\sqrt{2}}\sin\beta\mid C^{R*}_{2i}C^{L}_{1j}\mid\sin (\text{arg}(M_{2}\mu)$ ($C^{L, R}$ refer to unitary matrixes which diagonalize the chargino mass matrix);
whereas the CP-conserving Yukawa couplings in $a_{\mu}$ read as $\text{Re}(g^{P=A}_{\tilde{\chi}^{+}_{i}\tilde{\chi}^{-}_{j}})=-\frac{g}{\sqrt{2}}\sin\beta\mid C^{R*}_{2i}C^{L}_{1j}\mid$ 
and $\text{Re}(g^{S=h}_{\tilde{\chi}^{+}_{i}\tilde{\chi}^{-}_{j}})=-\frac{g}{\sqrt{2}}\sin\beta\mid C^{R*}_{1i}C^{L}_{2j}\mid$.

In Fig.\ref{chargino} we show the correlation between $a_{\mu}$ and $d_e$ from charginos.
For simplicity, we take nearly degenerate chargino mass spectrum $m_{\tilde{\chi}_{1}}\simeq m_{\tilde{\chi}_{2}}$.
In the {\it top} plot  $a_{\mu}$ is typically of order $\sim 10^{-12}$, which is about one order of magnitude larger than the result of stops.
Similar results hold in the {\it bottom} plot,  where the value of $d_{e}/e$ due to charginos are larger than those due to stops.
As previously shown in \cite{1612.08090}, 
the ACME-II limits can exclude the chargino mass below $10$ TeV for $\sin(\text{arg}(M_{2}\mu))$ of order unity.
Furthermore,  the {\it bottom} plot tells that 
the excursion limit is $\sim 1$ TeV for $\sin(\text{arg}(M_{2}\mu))$  of order $\sim 10^{-2}$,
and the magnitude of CP-violating phase $\sin(\text{arg}(M_{2}\mu))$ has to be of order $\sim 10^{-3}$ if sub-TeV charginos still survive
in the ACME-II limit.
Note, we have taken  $\text{sign}(\mu)=\text{sign}(M_{2})$ for our analysis. 
In the situation with an opposite assignment of signs, the numerical values of $a_{\mu}$ and $d_{e}$ both change, 
but their orders are still the same as those of identical sign assignment.

\begin{figure}
\centering
\begin{minipage}{8cm}
\centering
\includegraphics[width=8cm,height=4cm]{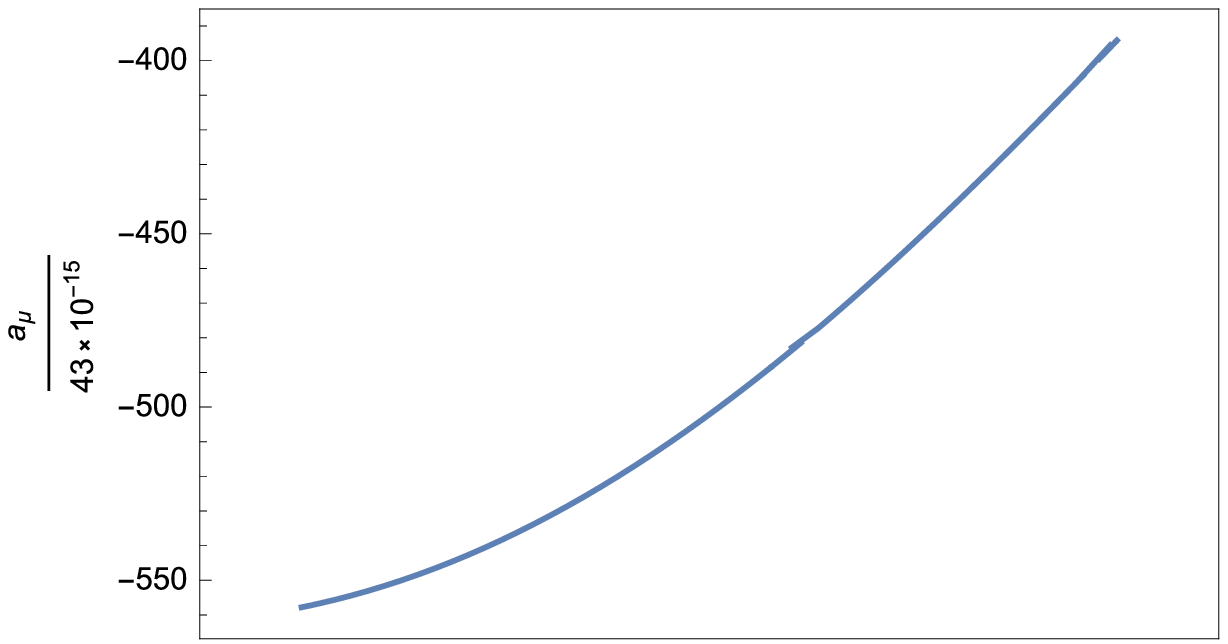}\\
\end{minipage}\\
~~~\centering
\begin{minipage}{8cm}
\centering
\includegraphics[width=7.85cm,height=4cm]{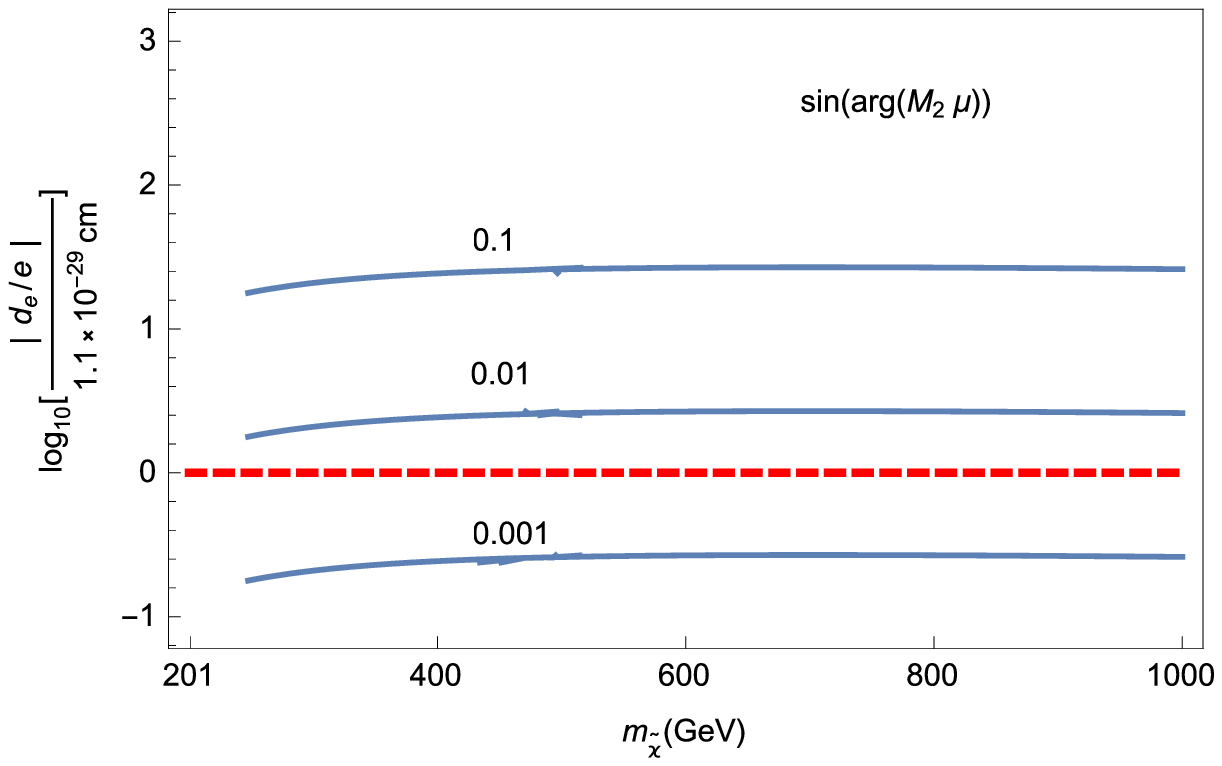}\\
\end{minipage}
 \caption{Correlated $a_{\mu}$ and $d_{e}$ from charginos.
 {\it Top}: the dependence of $a_{\mu}$ on nearly degenerate chargino mass $m_{\tilde{\chi}^{+}_{i}}$ for $m_{A}=1$ TeV.
 {\it Bottom}: the dependence of $d_e$ on chargino mass $m_{\tilde{\chi}}$ for  $\tan\beta=10$, $m_{A}=1$ TeV and different values of $\sin(\text{arg} (M_{2}\mu))=\{0.1,0.01,0.001\}$. The red line denotes the ACME-II limit.}
\label{chargino}
\end{figure}

Finally, we discuss the contributions to $d_e$ and $a_{\mu}$ due to the neutralino-chargino inner loop. 
According to the interaction 
\begin{eqnarray}{\label{ncl}}
\mathcal{L}\supset g\bar{\chi}^{-}_{i}\gamma^{\mu}[G_{ij}^{L}P_{L}+G_{ij}^{R}P_{R}]\chi^{0}_{j}W^{+}_{\mu}+\text{H.c},
\end{eqnarray}
where $G^{L}_{ij}=-C^{L*}_{1i}N_{2j}+\frac{1}{\sqrt{2}}C^{L*}_{2i}N_{4j}$ and $G^{R}_{ij}=-C^{R*}_{1i}N^{*}_{2j}-\frac{1}{\sqrt{2}}C^{R*}_{2i}N^{*}_{3j}$,
we have 
 \begin{eqnarray}{\label{nc}}
\text{Im}(c_{O})&=&-\frac{e\alpha}{\pi s^{2}_{W}} \text{Im}(G^{L}_{ij}G_{ij}^{R*})m_{\tilde{\chi}^{+}_{i}}m_{\tilde{\chi}^{0}_{j}}(1-x),\nonumber\\
\text{Re}(c_{O})&=&\frac{e\alpha}{\pi s^{2}_{W}} (\mid G_{ij}^{L}\mid^{2}-\mid G_{ij}^{R}\mid^{2})m^{2}_{\tilde{\chi}^{+}_{i}}(1-x)
\end{eqnarray}
and  $\Delta=[(1-x)m^{2}_{\tilde{\chi}^{+}_{i}}+xm^{2}_{\tilde{\chi}^{0}_{j}}]/x(1-x)$ \cite{0503055} in Eq.(\ref{analytic3}). 
Substituting them into Eq.(\ref{analytic3}) results in contributions to $d_e$ and $a_{\mu}$ as follows, 
\begin{eqnarray}{\label{neutralino-chargino}}
\frac{d_{e}/e}{1.1\times 10^{-29}\text{cm}}&=&-1.4\times 10^{3} \sum_{i,j}\int^{1}_{0} \frac{dx}{x} 
\text{Im}(G^{L}_{ij}G^{R*}_{ij})\frac{m_{\tilde{x}^{+}_{i}}}{100~\text{GeV}}\frac{m_{\tilde{x}^{0}_{j}}}{100~\text{GeV}}
j(0, \frac{1}{x}\frac{m^{2}_{\tilde{\chi}^{+}_{i}}}{m^{2}_{W}}+\frac{1}{1-x}\frac{m^{2}_{\tilde{\chi}^{0}_{j}}}{m^{2}_{W}})
\nonumber\\
\frac{a_{\mu}}{43\times10^{-10}}&=&-7.9\times 10^{-3}
\sum_{i,j}\int^{1}_{0} \frac{dx}{x}(\mid G_{ij}^{L}\mid^{2}-\mid G_{ij}^{R}\mid^{2})\frac{m^{2}_{\tilde{x}^{+}_{i}}}{\left(100~\text{GeV}\right)^{2}}
j(0, \frac{1}{x}\frac{m^{2}_{\tilde{\chi}^{+}_{i}}}{m^{2}_{W}}+\frac{1}{1-x}\frac{m^{2}_{\tilde{\chi}^{0}_{j}}}{m^{2}_{W}})\nonumber\\
\end{eqnarray}
wheres sum over $i=1-2$ and $j=1-4$. 
The result about $d_{e}$ in Eq.(\ref{neutralino-chargino}) agrees with that of the 3rd reference in \cite{electroweakino}.

\begin{figure}
\centering
\begin{minipage}{8cm}
\centering
~~\includegraphics[width=7.47cm,height=4cm]{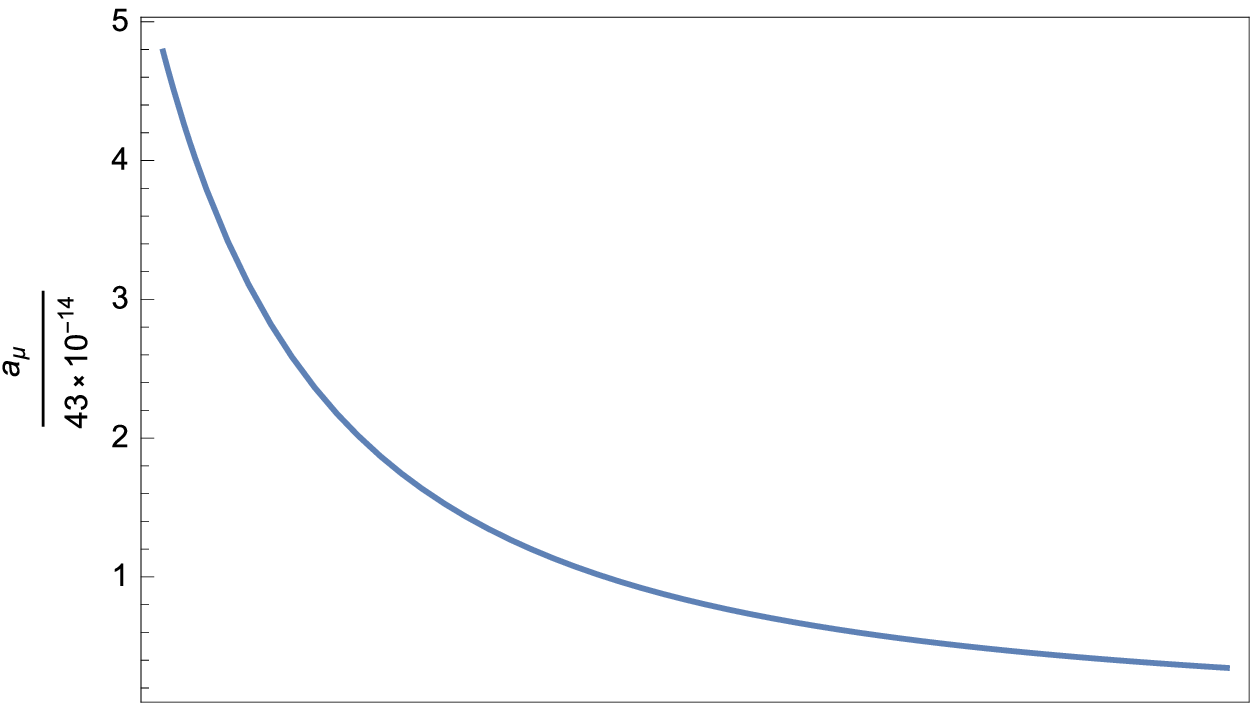}\\
\end{minipage}\\
\centering
\begin{minipage}{8cm}
\centering
\includegraphics[width=8cm,height=4cm]{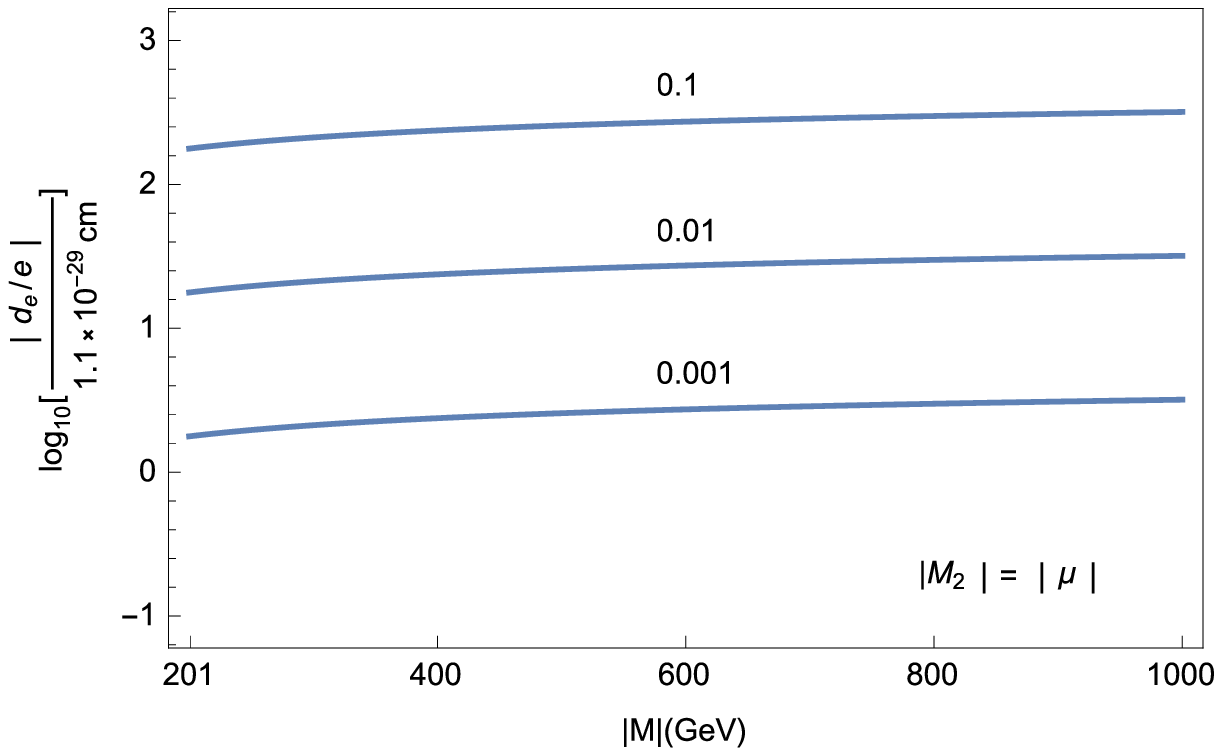}\\
\end{minipage}
 \caption{Correlated $a_{\mu}$ and $d_{e}$ from neutralinos and charginos with universal mass $M_{1,2}=\mu=M$.
 {\it Top}: the dependence of $a_{\mu}$ on $M$.
 {\it Bottom}: the dependence of $d_e$ on $M$ for different values of $\sin(\text{arg} (M_{2}\mu))=\{0.1,0.01,0.001\}$.}
\label{nc}
\end{figure}

From Eq.(\ref{neutralino-chargino}) it is clear that $a_{\mu}$ is sensitive to the mass spectrum of neutralinos and charginos.
When the mass splittings between neutralinos and charginos are small,
small $a_{\mu}$ is expected due to  small $\mid G_{ij}^{L}\mid^{2}-\mid G_{ij}^{R}\mid^{2}$ of order $\sim m^{2}_{Z}s^{2}_{W}/M^{2}$,
with $M$ referring to the universal neutralino and chargino mass. 
Fig.\ref{nc} shows the magnitudes of $a_{\mu}$ and $d_e$ in this situation. 
The {\it top} plot shows that $a_{\mu}$, which decreases as $M$ increases, is typically of order $\sim 10^{-13}$.
The {\it bottom} plot therein reveals that $d_e$, which is $\log$-dependent on $M$,  excesses the ACME-II limit 
unless the magnitude of $\sin(\text{arg} (M_{2}\mu))$ is smaller than $\sim 10^{-3}$.
Otherwise, $M$ beneath $1$ TeV has been excluded for large or moderate $\sin(\text{arg} (M_{2}\mu))$.
The reason for the $\log$-dependence of $d_e$ on $M$ is that the amplitude of  $G^{L}_{ij}G^{R*}_{ij}$ is of order unity,
which is not sensitive to the universal mass $M$ much heavier than $m_Z$.
Rather, a relatively large $a_{\mu}$ appears when there is large mass splittings between the masses of neutralino and charginos.
An optimistic estimate is that the enhancements on $a_{\mu}$ in Fig.\ref{nc} is about two orders of magnitude for 
$\mid G_{ij}^{L}\mid^{2}-\mid G_{ij}^{R}\mid^{2}$ of order unity,
which makes the contribution to $a_{\mu}$ due to neutralino-chargino loop dominates over the other two-loop effects.
However, we would like to mention that this case is strongly constrained by the limits from LHC Run 1 and 2 \cite{ncLHC}, 
since there is large missing energy with large mass splitting during the productions of either neutralinos or charginos at the LHC.

\section{Conclusions}
In this paper, we have explored the two-loop correlations between $a_{\mu}$ and $d_e$ due to either stops, charginos or neutralinos in natural SUSY. In the individual situation, $a_{\mu}$ can probe the superparticle masses,
whereas $d_e$ can determine the CP-violating phase tied to the same superparticle mass. 
Recent ACME-II limit has revealed that the magnitudes of CP-violating phases should be small, 
with the range of order $0.1$ to $0.001$ for sub-TeV stops and neutralinos/charginos, respectively.
On the other hand, the magnitudes of the correlated $a_{\mu}$ vary from $\sim 10^{-13}$ to $\sim 10^{-12}$ for stops and charginos, respectively,
which are beyond the scope of Brookhaven E821 ($\sim 10^{-9}$) but may be examined by 
future experiments on $a_{\mu}$ such as Fermilab E989 experiment.
Together with constraints such as flavor violation,
the EDM and MDM studied here offer us a complementary way to the direct detection on natural SUSY at future HL-LHC.
\\
\\
$\mathbf{Acknowledgments}$.
This work is supported by the National Natural Science Foundation of China (No. 11775039 and No. 11847301)
and the Fundamental Research Funds for the Central Universities at Chongqing University (No. cqu2017hbrc1B05).

\end{document}